\documentclass[a4paper,11pt]{article}

\usepackage{pos}
\usepackage{multicol,float, subcaption}
\usepackage{makecell}
\usepackage{graphicx}\usepackage{enumitem}
\usepackage{gensymb}
\usepackage{lipsum} 
\usepackage{lineno}

\title{TXS 0506+056 with Updated IceCube Data }

\ShortTitle{TXS 0506+056 with Updated IceCube Data }

\author{The IceCube Collaboration \\{\normalsize \normalfont(a complete list of authors can be found at the end of the proceedings)}\\}

\emailAdd{wluszczak@icecube.wisc.edu}

\abstract{

Past results from the IceCube Collaboration have suggested that the blazar TXS 0506+056 is a potential source of astrophysical neutrinos. However, in the years since there have been numerous updates to event processing and reconstruction, as well as improvements to the statistical methods used to search for astrophysical neutrino sources. These improvements in combination with additional years of data have resulted in the identification of NGC 1068 as a second neutrino source candidate. This talk will re-examine time-dependent neutrino emission from TXS 0506+056 using the most recent northern-sky data sample that was used in the analysis of NGC 1068. The results of using this updated data sample to obtain a significance and flux fit for the 2014 TXS 0506+056 "untriggered" neutrino flare are reported.

\vspace{4mm}
{\bfseries Corresponding authors:}
William Luszczak$^{*1,2}$\\
{$^{1}$ \itshape Dept. of Astronomy, Ohio State University, Columbus, OH 43210, USA}\\
{$^{2}$ \itshape Dept. of Physics and Center for Cosmology and Astro-Particle Physics, Ohio State University, Columbus, OH 43210, USA}\\[4mm]
$^*$ Presenter

\ConferenceLogo{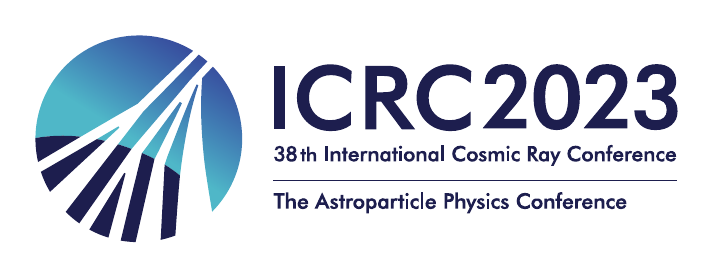}

\FullConference{The 38th International Cosmic Ray Conference (ICRC2023)\\ 26 July -- 3 August, 2023\\ Nagoya, Japan}
}

\begin{document}

\maketitle

\section{Introduction}\label{sec1}

TXS 0506+056 was originally identified as a potential neutrino emitter via a high energy IceCube event (IC 170922A) that triggered an alert that was subsequently followed up by a variety of other telescopes and experiments. A blazar (TXS 0506+056) was identified to be spatially coincident with the high energy IceCube event, at the level of $3 \sigma$ \cite{IceCube:TXS0506MMA}. 

Additionally, the IceCube collaboration performed a follow up analysis of archival neutrino data at the location of TXS 0506+056. This analysis identified a neutrino flare candidate several years prior to IC170922A, with a significance of $3.5 \sigma$. The alert event IC170922A does not contribute to the 2014 neutrino flare candidate, which obtains its significance from lower energy events earlier in the data sample. Notably, TXS 0506+056 did not display elevated gamma ray emission during the period of the neutrino flare candidate \cite{IceCube:TXSutf}. 

There have been numerous improvements in IceCube event processing, selection, and reconstruction over the past several years. A comparison of the most recent sample of track events with the sample used in the TXS 0506+056 untriggered flare search can be seen in table~\ref{tabel1}. Notable developments in data processing since the original TXS 0506+056 untriggered flare analysis include the implementation of pass 2 data processing, improved energy reconstruction, a BDT angular error estimator, and an improved likelihood that more accurately describes the per-event angular error.

Many of these improvements contributed to the identification of NGC 1068 as a candidate neutrino source with a significance of $4.2 \sigma $~\cite{IceCube:NGC1068}. Here, we explore the effect of using an updated IceCube event sample to perform an untriggered flare search at the location of TXS 0506+056. This contribution will use the following names in order to distinguish between the various relevant data samples used:

\begin{itemize}
    \item "Point Source Tracks v2" refers to the sample that was used for the untriggered flare search detailed in ref.~\cite{IceCube:TXSutf}. This analysis revealed a $3.5 \sigma$ neutrino flare candidate associated with TXS 0506+056 occurring in 2014. This sample was also used for the all-sky time-integrated analysis detailed in ref.~\cite{IceCube:7yrps}.
    \item "Northern Tracks v5" refers to the sample that was used for the time integrated searches presented in ref.~\cite{IceCube:NGC1068}. This sample is notable as many of the improvements that were introduced resulted in the identification of NGC 1068 as a neutrino source candidate at the level of $4.2 \sigma$.
\end{itemize}

\begin{table}
\begin{center}
\begin{tabular}{ |c|cc|} 
 \hline
   & Point Source Tracks v2 ~\cite{IceCube:TXSutf}  & Northern Tracks v5 ~\cite{IceCube:NGC1068}\\
 \hline \hline
 Pre-cuts & No & Yes\\ 
 BDT 1 & Selects tracks & Selects tracks \\
 BDT 2 & None & Rejects cascades \\
 Signal training set & Simulation & Simulation \\
 Background training set & Data & Simulation \\
 Angular error estimator & Paraboloid~\cite{paraboloid} & BDT \\
 Energy estimator & MuEX ~\cite{IceCube:ereco} & DNN \\ 
 DeepCore included? & Yes & No \\
 Livetime & 7 years (2008-2015) & 10 years (2011-2021) \\
 Declination Range & $-90^{\circ}<\delta<90^{\circ}$ & $-5^{\circ}<\delta<90^{\circ}$ \\
 Events & 711,878 & 794,301 \\
 \hline
\end{tabular}
\caption{\label{tabel1} A comparison of Point Source Tracks v2 and Northern Tracks v5. "Pre-cuts" refers to cuts on number of hit PMTs and track length that are performed prior to the BDT. }
\end{center}
\end{table}

\section{Untriggered Flare Search at the Location of TXS 0506+056 Using Northern Tracks v5}
We attempt to match the methodology of the original TXS 0506+056 untriggered flare analysis as closely as possible, save for the updated data and analysis methods. The data sample used is Northern Tracks v5, previously used for the time integrated analysis documented in ref.~\cite{IceCube:NGC1068}. The data sample is divided into several different seasons as shown in table~\ref{table3}. This slightly differs from the original divisions shown in table 1 of ref.~\cite{IceCube:TXSutf} as seasons prior to May 13, 2011 have been excluded, due to Northern Tracks v5 processing excluding these seasons from the data sample. The most recent search period ends on October 31, 2017, and does not include data beyond that point in order to replicate the original analysis \cite{IceCube:TXSutf}. Existing data beyond this date is excluded, as the goal of this study is to examine the effect of the new data sample and methods on the 2014 neutrino flare candidate. To this end, the results in table~\ref{table3} compare the p-value associated with the best-fit flare without the livetime trial factor of $9.5/3$ used in ref.~\cite{IceCube:TXSutf}. 

The test statistic is constructed from a likelihood ratio comparing the best-fit box shaped flare with the null hypothesis of no neutrino emission. This follows the generic time-dependent likelihood often used in transient neutrino point source analyses~\cite{Braun_2010}, with the exception that the spatial and energy components of the likelihood utilize the KDE-based approach used in the most-recent Northern Tracks v5 time-integrated analysis. The full likelihood can be written as:

\begin{equation}
    \mathcal{L}(n_{s}, \gamma) = \prod_{i=1}^{N} \left( \frac{n_{s}}{N}S_{i} + (1-\frac{n_{s}}{N})B_{i} \right).
\end{equation}

\noindent
$N$ represents the total number of events in the sample, $S_i$ and $B_i$ are PDFs describing the probability that the $i$th event is produced by signal and background, respectively, and $n_s$ and $\gamma$ are the fitted number of signal events and flux spectral index. Differing from the original analysis, the signal PDF is now expressed as:

\begin{equation}
    S(E_{\mathrm{reco}}, \Psi, \sigma | \delta_{\mathrm{src}}, \gamma) = f_s(\Psi|E_{\mathrm{reco}}, \sigma, \gamma) \times f_s(E_{\mathrm{reco}}, \sigma|\delta_{\mathrm{src}}, \gamma) \times \tau(t, T_{\mathrm{start}}, T_{\mathrm{stop}}),
\end{equation}

\noindent
where $E_{\mathrm{reco}}$ is the reconstructed event energy, $\Psi$ is the angular separation between a particular event and the source candidate, $\delta_{src}$ is the declination of the source, and $\sigma$ is the per-event angular error. The first two terms (describing the spatial and energy components) are now jointly constructed from KDEs, instead of the previously used product of independent spatial and energy components. This represents an improvement over the previous method, which in many cases misrepresents the point spread function. The temporal PDF $\tau(t, T_{\mathrm{start}}, T_{\mathrm{stop}})$ is a box PDF:

\begin{equation}
    \tau(t, T_{\mathrm{start}}, T_{\mathrm{stop}}) = 
    \begin{cases}
        \frac{1}{T_{\mathrm{stop}}-T_{\mathrm{start}}} & \text{if \; $T_{\mathrm{start}}<t<T_{\mathrm{stop}}$} \\
        0 & \text{otherwise}, 
    \end{cases}
\end{equation}

\noindent
where $T_{start}$ and $T_{stop}$ describe the start and stop times of the flare candidate. While ref.~\cite{IceCube:TXSutf} additionally tests the hypothesis of a gaussian-shaped flare, this modification is not presented here. 

The likelihood is then maximized to find best fit values for $T_{start}$, $T_{stop}$, $n_s$, and $\gamma$ at the location of TXS 0506+056. Note that the flare parameters are fit by the likelihood, not decided a-priori or according to an external lightcurve, hence the descriptor "untriggered". To calculate a significance, the test statistic is compared to a distribution of test statistics representing the null hypothesis. This null hypothesis distribution is obtained by calculating the test statistic for trials with the events rotated by a random amount in right ascension, effectively destroying any potential clustering present.

\begin{table}
\begin{center}
\begin{tabular}{ |c|cc|} 
 \hline
 Season & Start  & End\\
 \hline \hline
 IC86a &  2011 May 13  &  2012 May 16 \\
 IC86b &  2012 May 16  &  2015 May 18 \\
 IC86c &  2015 May 18  &  2017 Oct 31 \\
 \hline
\end{tabular}
\caption{\label{table3} The seasons into which IceCube data has been divided for the purposes of an untriggered flare search. Like the analysis presented in ref.~\cite{IceCube:TXSutf}, seasons are searched for flare candidates independently, and flare candidates must be fully contained within one season. IceCube seasons prior to May 13, 2011 are not used, as the Northern Tracks v5 event selection has not been developed to function for seasons with only a partial ($<86$ strings) detector configuration. }
\end{center}
\end{table}

\section{Expected Performance of an Untriggered Flare Search with Updated Data}

\begin{figure}[t!]
    \centering
    \caption{Performance comparison for an untriggered flare search}
    \begin{subfigure}[t]{0.49\textwidth}
        \centering
        \includegraphics[width=0.99\textwidth]{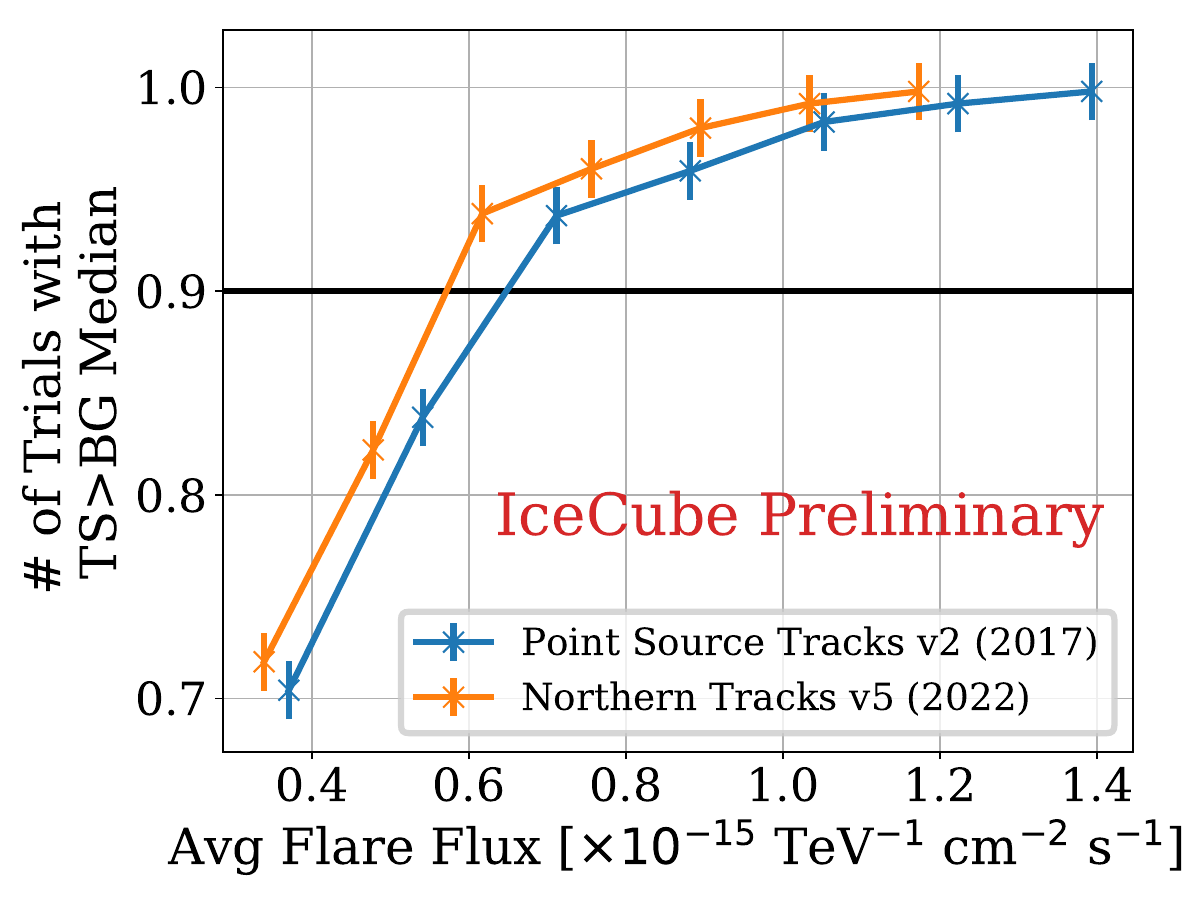}
        \subcaption{Fraction of trials with an untriggered neutrino flare TS greater than the background median across a range of injected flare fluxes. All injected flares are simulated with a spectral index of $\gamma=2.2$ and a duration of $\Delta T =$ 158 days, in accordance with the best-fit values obtained in the previous untriggered flare search~\cite{IceCube:TXSutf}.} 
    \end{subfigure}
    ~
    \begin{subfigure}[t]{0.49\textwidth}
        \centering
        \includegraphics[width=0.99\textwidth]{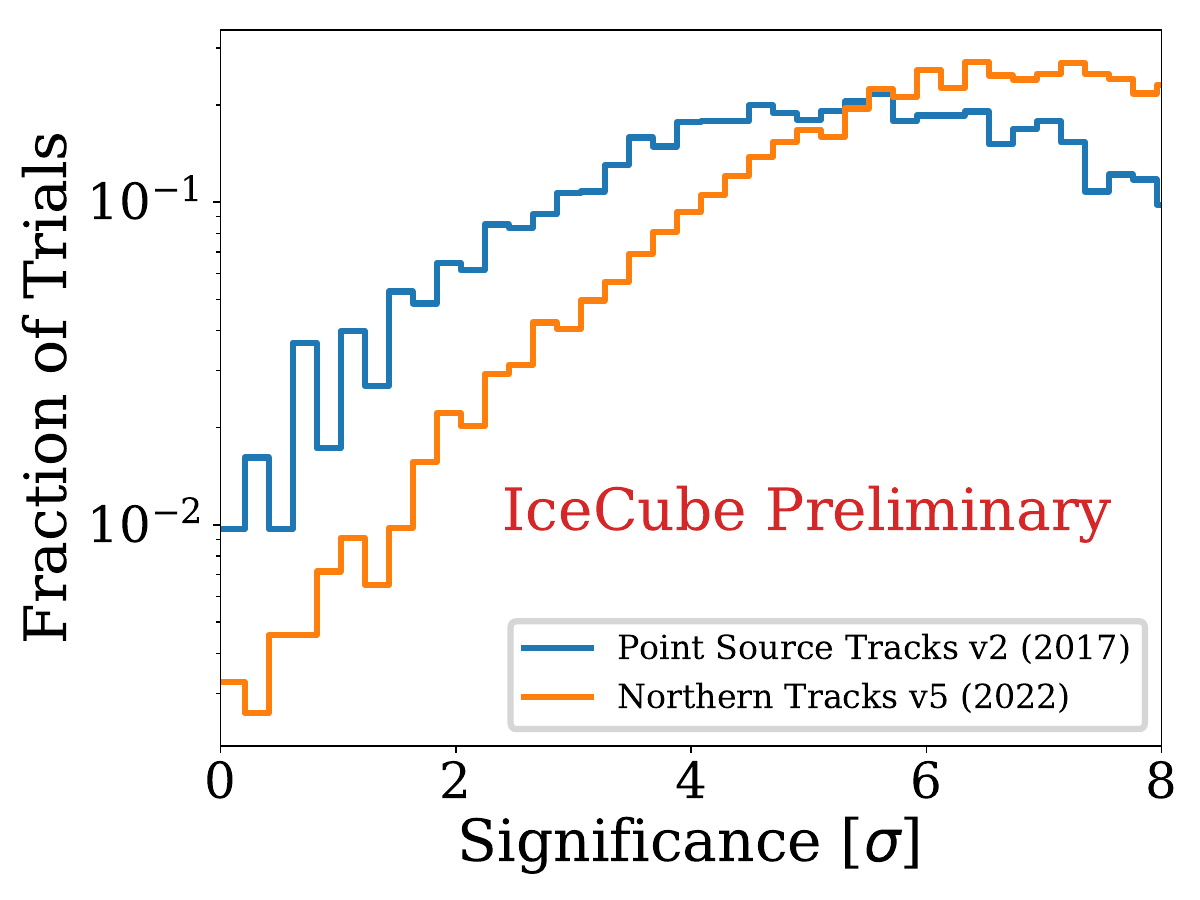}
        \subcaption{Distributions for the expected significance of an untriggered flare fit for MC simulations of a TXS 0506+056-strength flare using the original 2017 data sample (blue), and the updated 2022 data sample (orange).}
    \end{subfigure}
    \label{fig:fig1}
\end{figure}

 The efficiency and expected significance of a flare with flux equivalent to the best-fit flare flux from ref.~\cite{IceCube:TXSutf} are used as metrics to evaluate the performance of an untriggered flare search. Figure~\ref{fig:fig1} shows efficiency curves for simulated 158 day flares with spectral index $\gamma=2.2$ and various flux normalizations. The newer Northern Tracks v5 sample represents an improvement over older data samples, as the analysis becomes more efficient at finding flares with lower flux normalizations. 
 
The expected significance of the TXS 0506+056 2014 neutrino flare candidate is  calculated as follows: Simulated flares with a duration of 158 days, flux normalization at 100 TeV of $\Phi_{100 \mathrm{TeV}} = 1.6 \times 10^{-15}$ TeV$^{-1}$ cm$^{-2}$ s$^{-1}$ and spectral index $\gamma=2.2$ were injected into the sample, and the untriggered flare search described above was performed. Significances for each trial were calculated and are shown in figure~\ref{fig:fig1}. In both Point Source Tracks v2 and Northern Tracks v5, the distributions of expected significance are quite wide, however on average TXS-0506+056--like neutrino flare candidates are expected to have higher significance in an untriggered flare search when using Northern Tracks v5.

\section{Results}
The results of performing an untriggered flare search at the location of TXS 0506+056 using Northern Tracks v5 are shown table~\ref{table2}. For comparison, the results from the original analysis using Point Source Tracks v2 are also shown. 

Notably, Northern Tracks v5 fits a smaller flux normalization than Point Source Tracks v2, along with a correspondingly lower significance. Further investigation reveled the cause to be two partially contained cascade events present in Point Source Tracks v2 that have since been removed in newer data samples. These events fail to pass a track length pre-cut that is present in the selection criteri for Northern Tracks v5 (tracks must be longer than 200 meters), but not present in the criteria for Point Source Tracks v2. Though these cascade-like events have been removed from newer tracks-based samples, they were present in Point Source Tracks v2, and thus contributed to the result reported in ref.~\cite{IceCube:TXSutf}. The reconstructions used in Point Source Tracks v2 assume that events are track-like, and can return inaccurate values when used on events with a cascade geometry. If reconstructed as cascades, neither of these events point in the direction of TXS 0506+056. As Northern Tracks v5 does not contain these events, the fitted flux and significance are lower when using this event sample. The two cascade-like events have also been removed from the IceCube public-release point source data, with a similar effect on the untriggered flare results noted in ref. ~\cite{https://doi.org/10.21234/cpkq-k003}.

Figure~\ref{fig:fig3} shows the events near TXS 0506+056 that contribute most to the 2014 flare, both in Point Source Tracks v2 and Northern Tracks v5, with the two cascade events that are not present in Northern Tracks v5 shown as dashed circles. The effect of removing the two cascade-like events from the sample can be seen in the spatial scans shown in figure~\ref{fig:fig5}. When using Northern Tracks v5, the hotspot associated with TXS 0506+056 in neutrinos during the 2014 flare candidate period has shifted, and the significance associated with the maximum has also decreased.

It is important to note that the existence of these cascade-like events in the Point Source v2 data sample does not invalidate the originally calculated significance. Similar events are present in simulation as well, and are accounted for by the analysis procedures. Additionally, correlation studies were performed: events were injected into Point Source Tracks v2 according to the original best-fit flare flux. Any events not present in the newer samples were removed and a significance using the same events in the newer samples was calculated. A drop in significance of the scale seen here occurred in approximately 8\% of the correlation study trials, indicating consistency between the newer and older data samples.

\begin{table}
\begin{center}
\begin{tabular}{ |c|ccccc|} 
 \hline
 Sample & \makecell{$p$ \\ (pre-trial)} & \makecell{Best-Fit Flux \\ ($\times 10^{-15}$ TeV$^{-1}$ cm$^{-2}$ s$^{-1}$)} & $\gamma$ & \makecell{$T_{start}$\\ (MJD)} & \makecell{$T_{stop}$\\(MJD)} \\
 \hline \hline
Point Source Tracks v2 & 7e-5 & 1.6 & 2.2 & 56937.81 & 57096.22\\
Northern Tracks v5 & 1e-3 & 0.76 & 2.2 & 56927.86 & 57091.33\\
 \hline
\end{tabular}
\caption{\label{table2} The results of an untriggered flare search at the location of TXS 0506+056 using older (Point Source Tracks v2) and newer (Northern Tracks v5) data. The best-fit flux of the 2014 flare candidate is approximately half as large when using the newer data sample of Northern Tracks v5, with a correspondingly lowered significance. Here, "pre-trial" excludes any factor accounting for dividing up the livetime into seasons that are then independently searched. This factor was 9.5/3 in the
original follow-up analysis.  }
\end{center}
\end{table}

\begin{figure}[h!]
    \centering
    \caption{Untriggered Flare Search Results}
    \begin{subfigure}[t]{0.49\textwidth}
        \centering
        \includegraphics[width=0.99\textwidth]{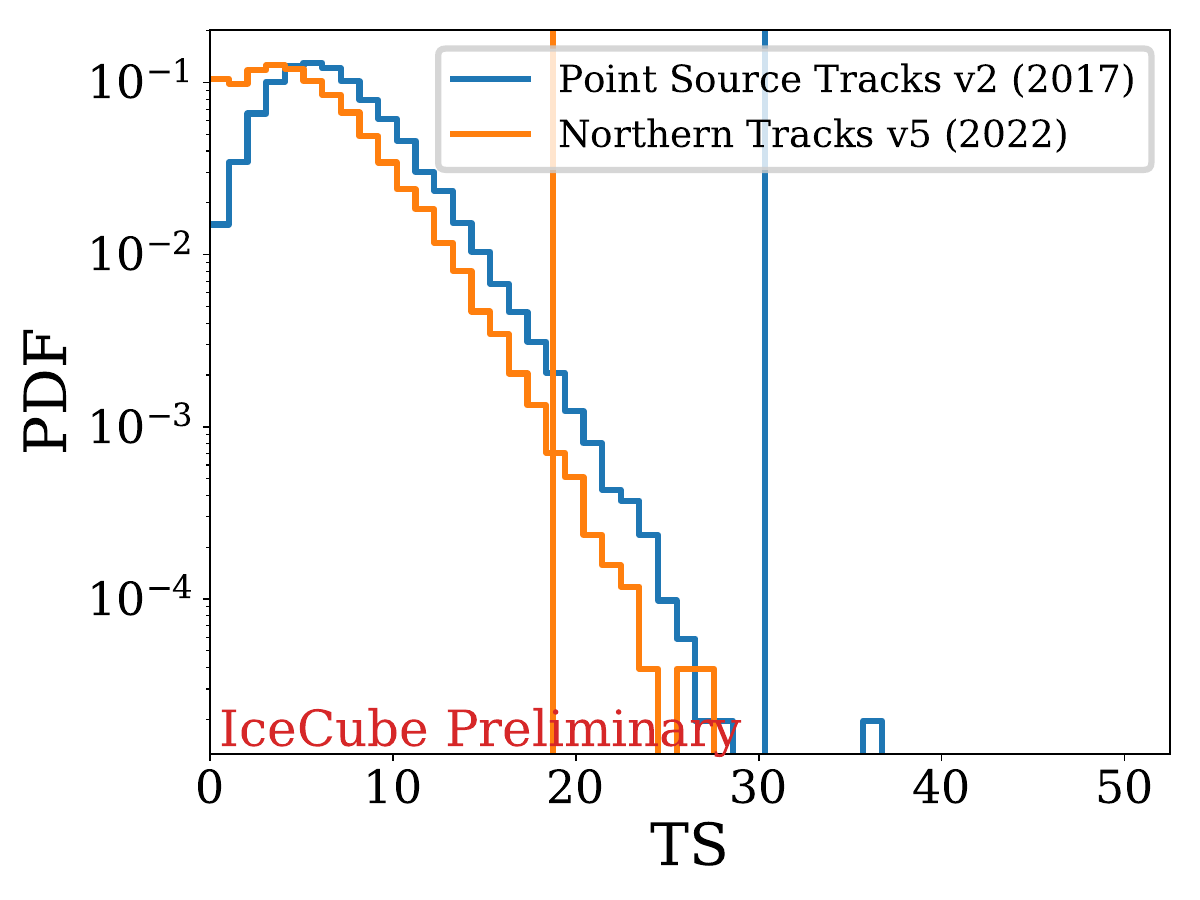}
        \subcaption{Test statistic distributions for the null hypothesis (no injected flare) for both the 2017 data sample and the updated 2022 data sample. The measured untriggered flare test statistic for each sample is shown as a vertical line}
    \end{subfigure}
    ~
    \begin{subfigure}[t]{0.49\textwidth}
        \centering
        \includegraphics[width=0.99\textwidth]{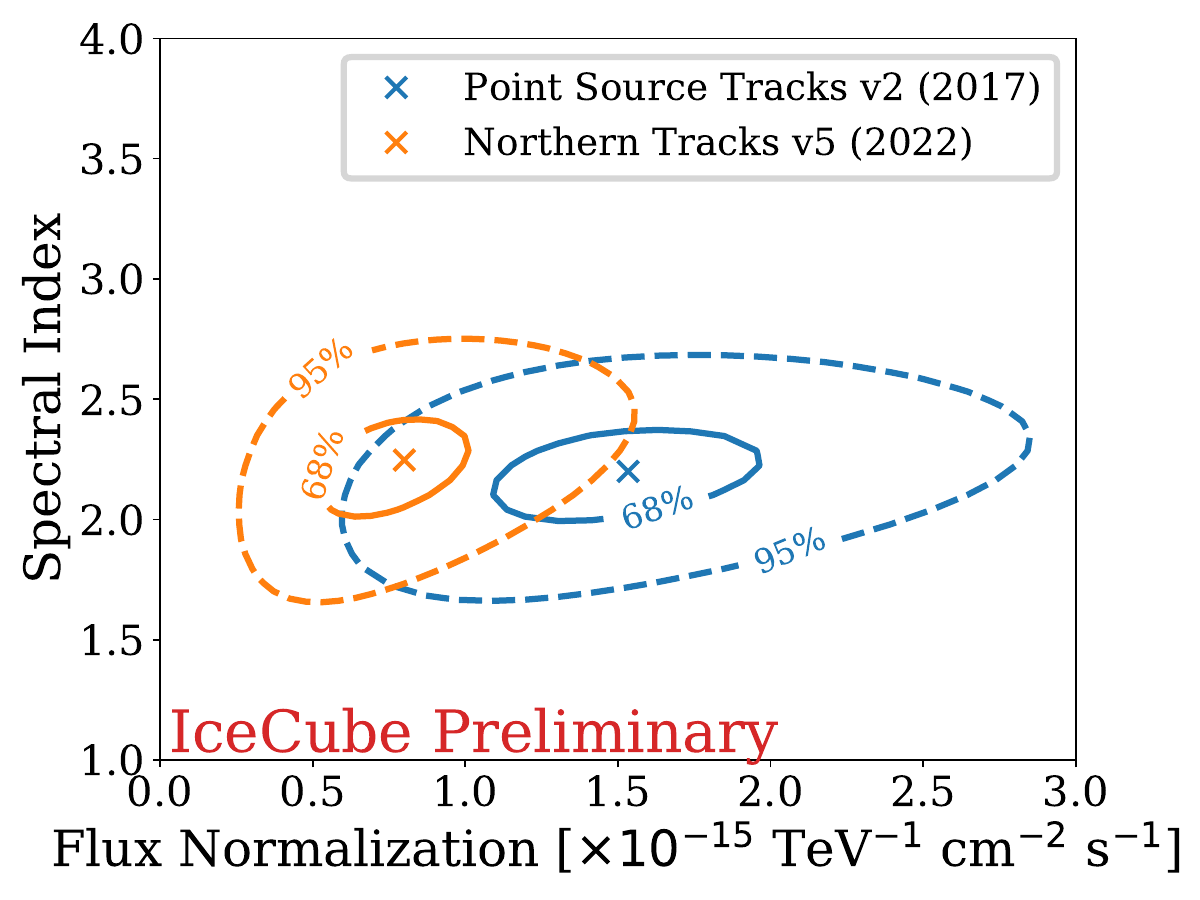}
        \subcaption{Best fit values for the 2014 TXS 0506+056 untriggered neutrino flare using the original 2017 data sample (blue) and the updated 2022 sample (orange). Solid and dashed lines correspond to contours of 68\% and 95\% containment, respectively }
    \end{subfigure}
    \label{fig:fig2}
\end{figure}

\begin{figure}[h!]
    \centering
    \begin{subfigure}[t]{0.49\textwidth}
        \centering
        \includegraphics[width=0.99\textwidth]{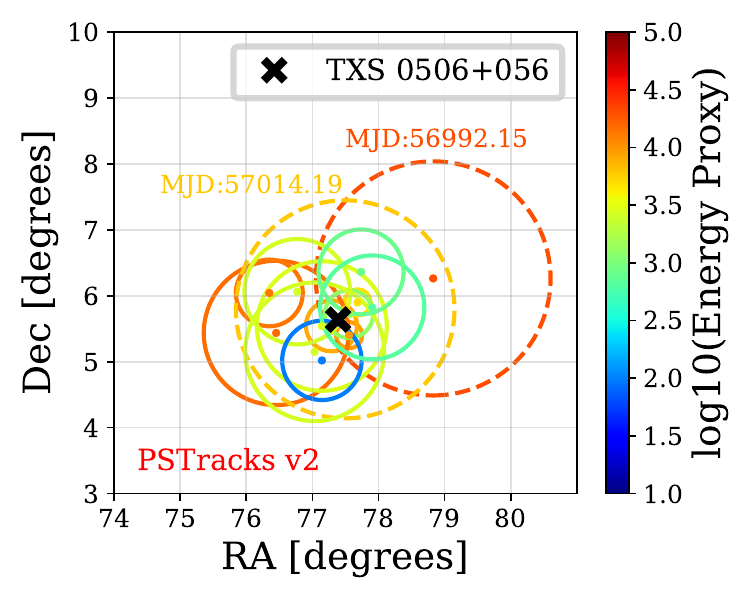}
    \end{subfigure}
    ~
    \begin{subfigure}[t]{0.49\textwidth}
        \centering
        \includegraphics[width=0.99\textwidth]{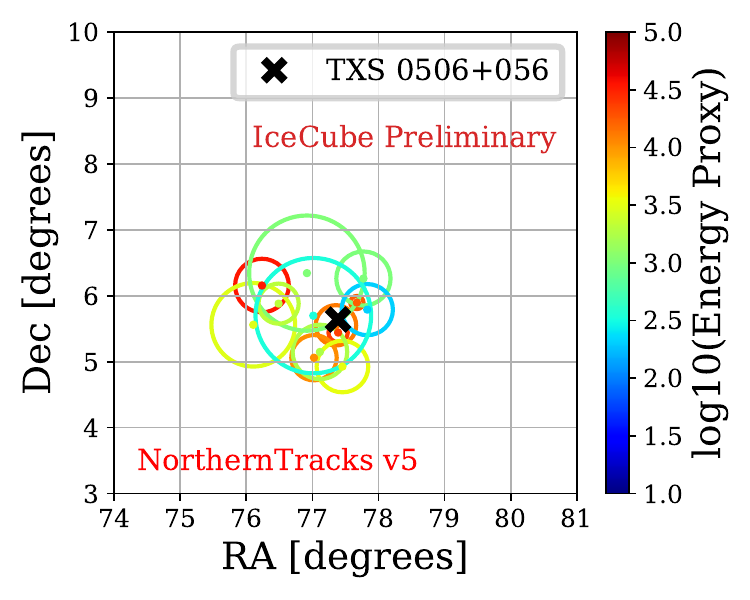}
    \end{subfigure}
    \caption{The locations of the 13 events that contribute most to the 2014 flare in Point Source Tracks v2 (left, taken from ref.~\cite{https://doi.org/10.21234/cpkq-k003}) and Northern Tracks v5 (right). The circles represent the 1-sigma angular error regions for each event, and the color corresponds to the event energy proxy.}
    \label{fig:fig3}
\end{figure}

\begin{figure}[h!]
    \centering
    \begin{subfigure}[t]{0.49\textwidth}
        \centering
        \includegraphics[width=0.99\textwidth]{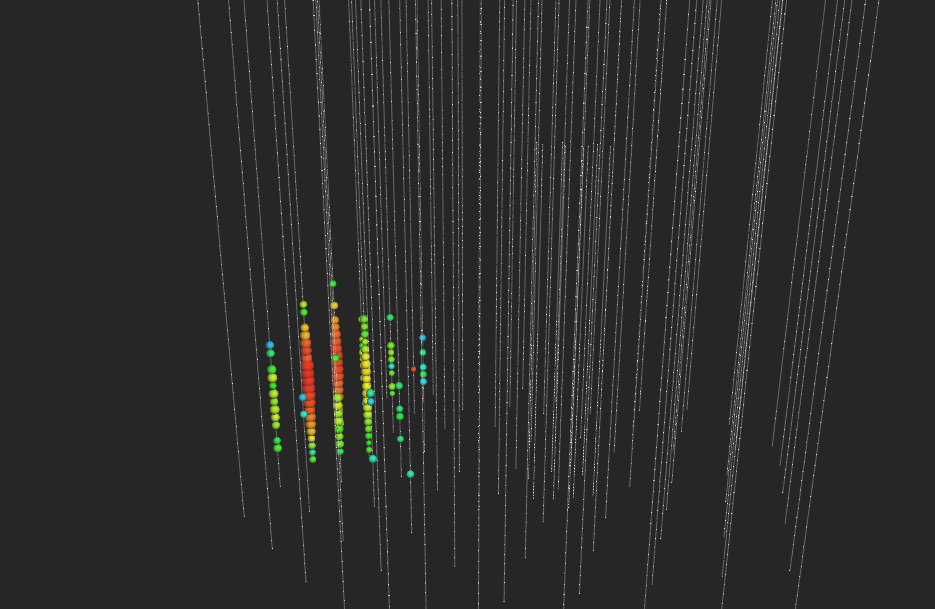}
    \end{subfigure}
    ~
    \begin{subfigure}[t]{0.49\textwidth}
        \centering
        \includegraphics[width=0.99\textwidth]{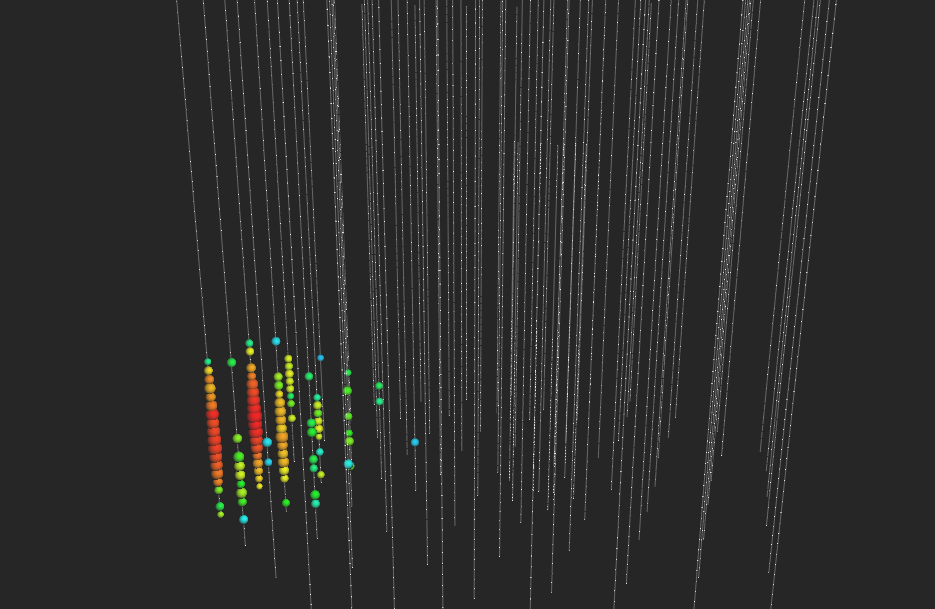}
    \end{subfigure}
    \caption{Two corner-clipping cascade events present in the 2017 data sample that contributed to the original result presented in ref~\cite{IceCube:TXSutf}. While originally reconstructed as tracks in Point Source Tracks v2, if these events are reconstructed as cascades they do not point in the direction of TXS 0506+056. Newer tracks samples like Northern Tracks v5 remove these events entirely. }
    \label{fig:fig4}
\end{figure}

\begin{figure}[h!]
    \centering
    \begin{subfigure}[t]{0.49\textwidth}
        \centering
        \includegraphics[width=0.99\textwidth]{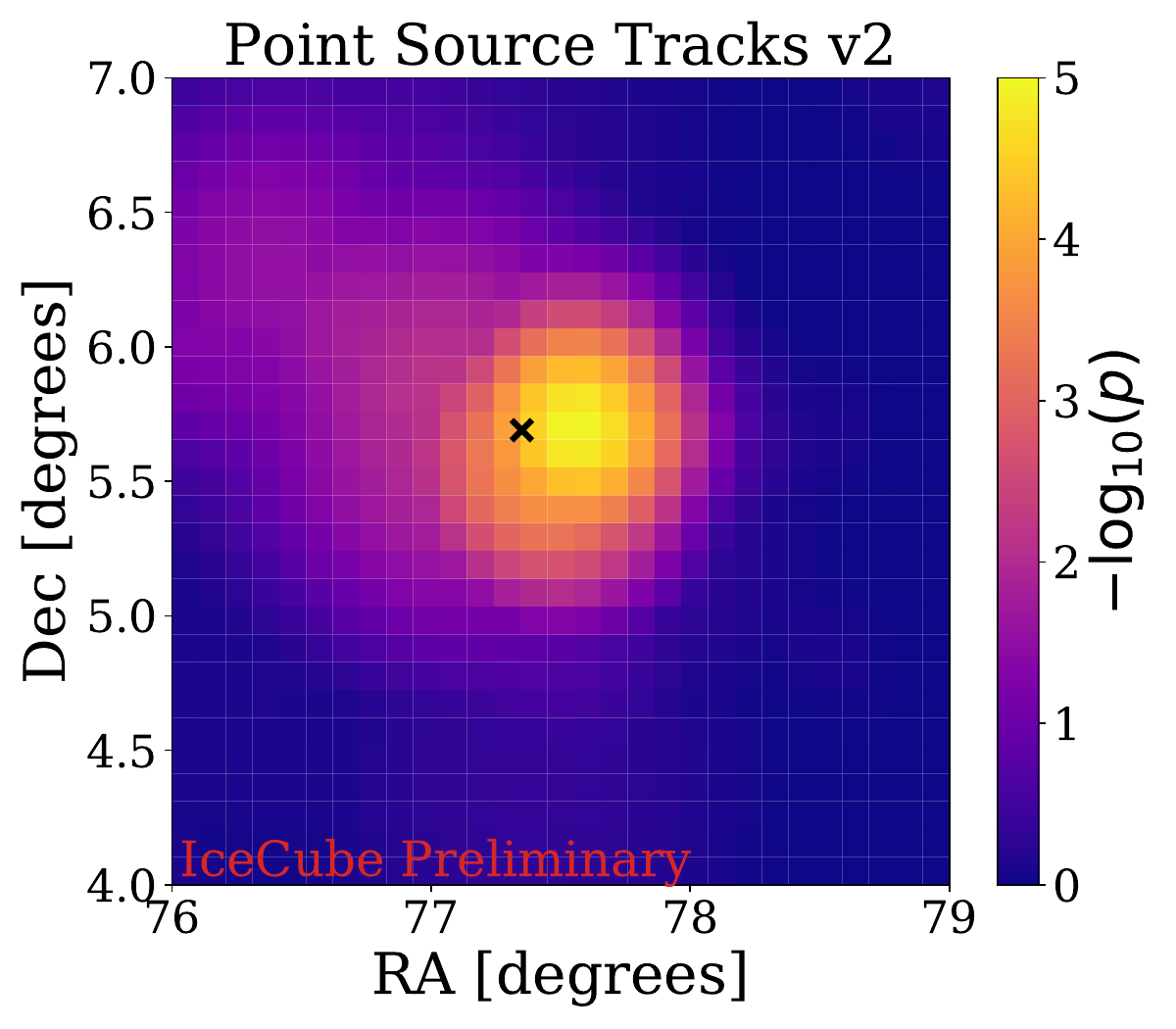}
    \end{subfigure}
    ~
    \begin{subfigure}[t]{0.49\textwidth}
        \centering
        \includegraphics[width=0.99\textwidth]{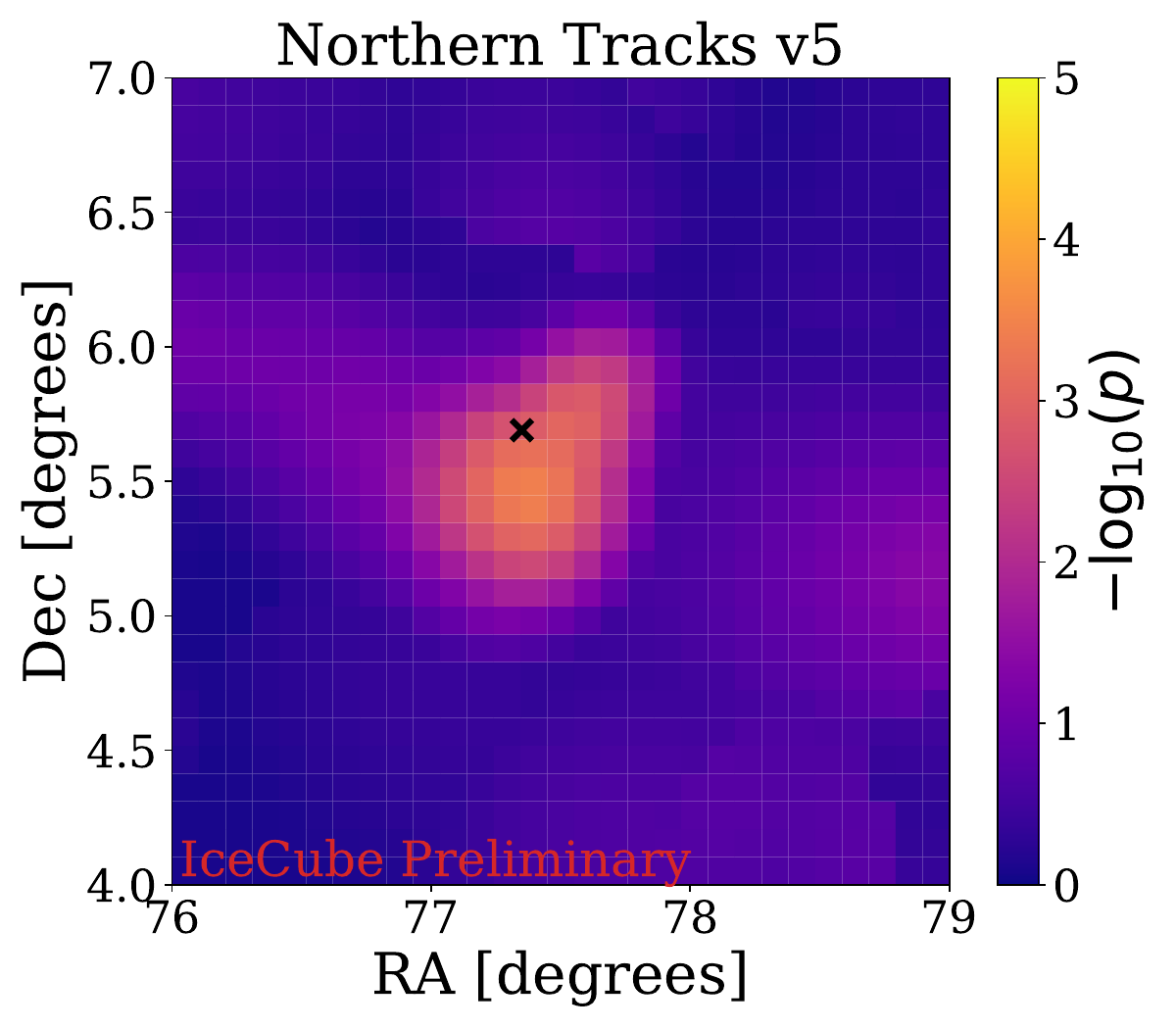}
    \end{subfigure}
    \caption{Spatial scans of the 2014 neutrino flare candidate near the location of TXS 0506+056, for both the original 2017 data sample (left) and the updated 2022 sample (right). In both plots, the best-fit flare start, stop, and spectral index are re-evaluated at each location. The location of TXS 0506+056 is marked with a black cross.}
    \label{fig:fig5}
\end{figure}

\section{Conclusion}\label{sec3}

This document details the effect of performing an untriggered flare search at the location of TXS 0506+056 using updated IceCube data. We find that when using the newer IceCube data sample of Northern Tracks v5, a smaller flux normalization (and correspondingly smaller significance) is obtained. This is primarily due to improved event selection techniques removing two cascade-like events from the newer IceCube track event data samples. These results are not inconsistent with previously published results, with correlation studies revealing similar results occur approximately 8\% of the time. It should be noted that one-zone models of neutrino emission from TXS 0506+056 struggle to produce a neutrino flare of the magnitude that was observed in 2014 while simultaneously fitting the neutrino data~\cite{Petropoulou_2020}. While fits with newer data producing a lower 2014 neutrino flare candidate flux lessen this tension, they do not alleviate it entirely. 
\vspace{2cm}

\bibliographystyle{ICRC}
\bibliography{references}

%

\clearpage

\section*{Full Author List: IceCube Collaboration}

\scriptsize
\noindent
R. Abbasi$^{17}$,
M. Ackermann$^{63}$,
J. Adams$^{18}$,
S. K. Agarwalla$^{40,\: 64}$,
J. A. Aguilar$^{12}$,
M. Ahlers$^{22}$,
J.M. Alameddine$^{23}$,
N. M. Amin$^{44}$,
K. Andeen$^{42}$,
G. Anton$^{26}$
% ...
\\
\\
% ...
$^{12}$ Universit{\'e} Libre de Bruxelles, Science Faculty CP230, B-1050 Brussels, Belgium \\
% ...
$^{17}$ Department of Physics, Loyola University Chicago, Chicago, IL 60660, USA \\
$^{18}$ Dept. of Physics and Astronomy, University of Canterbury, Private Bag 4800, Christchurch, New Zealand \\
% ...
$^{22}$ Niels Bohr Institute, University of Copenhagen, DK-2100 Copenhagen, Denmark \\
$^{23}$ Dept. of Physics, TU Dortmund University, D-44221 Dortmund, Germany \\
% ...
$^{26}$ Erlangen Centre for Astroparticle Physics, Friedrich-Alexander-Universit{\"a}t Erlangen-N{\"u}rnberg, D-91058 Erlangen, Germany \\
% ...
$^{40}$ Dept. of Physics and Wisconsin IceCube Particle Astrophysics Center, University of Wisconsin{\textendash}Madison, Madison, WI 53706, USA \\
% ...
$^{42}$ Department of Physics, Marquette University, Milwaukee, WI, 53201, USA \\
% ...
$^{44}$ Bartol Research Institute and Dept. of Physics and Astronomy, University of Delaware, Newark, DE 19716, USA \\
% ...
$^{63}$ Deutsches Elektronen-Synchrotron DESY, Platanenallee 6, 15738 Zeuthen, Germany  \\
$^{64}$ Institute of Physics, Sachivalaya Marg, Sainik School Post, Bhubaneswar 751005, India \\
% ...

\subsection*{Acknowledgements}

\noindent
The authors gratefully acknowledge the support from the following agencies and institutions:
USA {\textendash} U.S. National Science Foundation-Office of Polar Programs,
U.S. National Science Foundation-Physics Division,
U.S. National Science Foundation-EPSCoR,
Wisconsin Alumni Research Foundation,
Center for High Throughput Computing (CHTC) at the University of Wisconsin{\textendash}Madison,
Open Science Grid (OSG),
Advanced Cyberinfrastructure Coordination Ecosystem: Services {\&} Support (ACCESS),
Frontera computing project at the Texas Advanced Computing Center,
U.S. Department of Energy-National Energy Research Scientific Computing Center,
Particle astrophysics research computing center at the University of Maryland,
Institute for Cyber-Enabled Research at Michigan State University,
and Astroparticle physics computational facility at Marquette University;
Belgium {\textendash} Funds for Scientific Research (FRS-FNRS and FWO),
FWO Odysseus and Big Science programmes,
and Belgian Federal Science Policy Office (Belspo);
Germany {\textendash} Bundesministerium f{\"u}r Bildung und Forschung (BMBF),
Deutsche Forschungsgemeinschaft (DFG),
Helmholtz Alliance for Astroparticle Physics (HAP),
Initiative and Networking Fund of the Helmholtz Association,
Deutsches Elektronen Synchrotron (DESY),
and High Performance Computing cluster of the RWTH Aachen;
Sweden {\textendash} Swedish Research Council,
Swedish Polar Research Secretariat,
Swedish National Infrastructure for Computing (SNIC),
and Knut and Alice Wallenberg Foundation;
European Union {\textendash} EGI Advanced Computing for research;
Australia {\textendash} Australian Research Council;
Canada {\textendash} Natural Sciences and Engineering Research Council of Canada,
Calcul Qu{\'e}bec, Compute Ontario, Canada Foundation for Innovation, WestGrid, and Compute Canada;
Denmark {\textendash} Villum Fonden, Carlsberg Foundation, and European Commission;
New Zealand {\textendash} Marsden Fund;
Japan {\textendash} Japan Society for Promotion of Science (JSPS)
and Institute for Global Prominent Research (IGPR) of Chiba University;
Korea {\textendash} National Research Foundation of Korea (NRF);
Switzerland {\textendash} Swiss National Science Foundation (SNSF);
United Kingdom {\textendash} Department of Physics, University of Oxford.

\end{document}